\documentclass[final,5p,times,twocolumn]{elsarticle}

\usepackage{graphicx}

\usepackage{amssymb}

\usepackage{xspace}
\usepackage{textcomp}

\newcommand{\mte}{$\mu^+ \rightarrow e^+e^-e^+$\xspace}

\newcommand{\mtenunu}{$\mu^+ \rightarrow e^+e^-e^+\nu\bar{\nu}$\xspace}

\newcommand{\ord}{{\cal O}}

\journal{Nuclear Instruments and Methods A}

\begin{document}

\begin{frontmatter}



\title{A Tracker for the Mu3e Experiment based on High-Voltage Monolithic Active Pixel Sensors}


\author[A]{Niklaus Berger}
\author[A]{Heiko Augustin} 
\author[A]{Sebastian Bachmann} 
\author[A]{Moritz Kiehn}
\author[B]{Ivan Peri\'c}
\author[A]{Ann-Kathrin Perrevoort} 
\author[A]{Raphael Philipp} 
\author[A]{Andr\'e Sch\"oning}
\author[A]{Kevin Stumpf}
\author[A]{Dirk Wiedner}
\author[A]{Bernd Windelband}
\author[A]{Marco Zimmermann}
\address[A]{Physikalisches Institut, Heidelberg University, Heidelberg, Germany}
\address[B]{Zentralinstitut f\"ur technische Informatik, Heidelberg University, Mannheim, Germany}

\begin{abstract}

The Mu3e experiment searches for the lepton flavour violating decay $\mu^+\rightarrow e^+e^-e^+$, aiming for a branching fraction sensitivity of $10^{-16}$. 
This requires an excellent momentum resolution for low energy electrons, high rate capability and a large acceptance. 
In order to minimize multiple scattering, the amount of material has to be as small as possible. 
These challenges can be met with a tracker built from high-voltage monolithic active pixel sensors (HV-MAPS), which can be thinned to 50~\textmu m and which incorporate the complete read-out electronics on the sensor chip. 
To further minimise material, the sensors are supported by a mechanical structure built from 25~$\mu$m thick Kapton foil and cooled with gaseous helium. 
\end{abstract}


\end{frontmatter}


\section{Motivation}

In the standard model of elementary particles (SM), lepton flavour is a conserved quantity. 
In the neutrino sector, lepton flavour violation (LFV) has however been observed in the form of neutrino mixing. 
Consequently, lepton flavour symmetry is a broken symmetry, the standard model has to be adapted to incorporate massive neutrinos and LFV is also expected in the charged lepton sector, but has so far not been observed. 
The exact mechanism and size of LFV being unknown, its study is of large interest, as it is linked to neutrino mass generation, CP
violation and new physics beyond the SM. 
In fact, even in a Standard Model extended with massive neutrinos, the branching fraction for decays like \mte is suppressed to unobservable levels of $\ord(10^{-50})$; an observation would thus be an unequivocal sign for new physics.

\begin{figure}[b!] 
\centering 
\includegraphics[width=0.9\columnwidth,keepaspectratio]{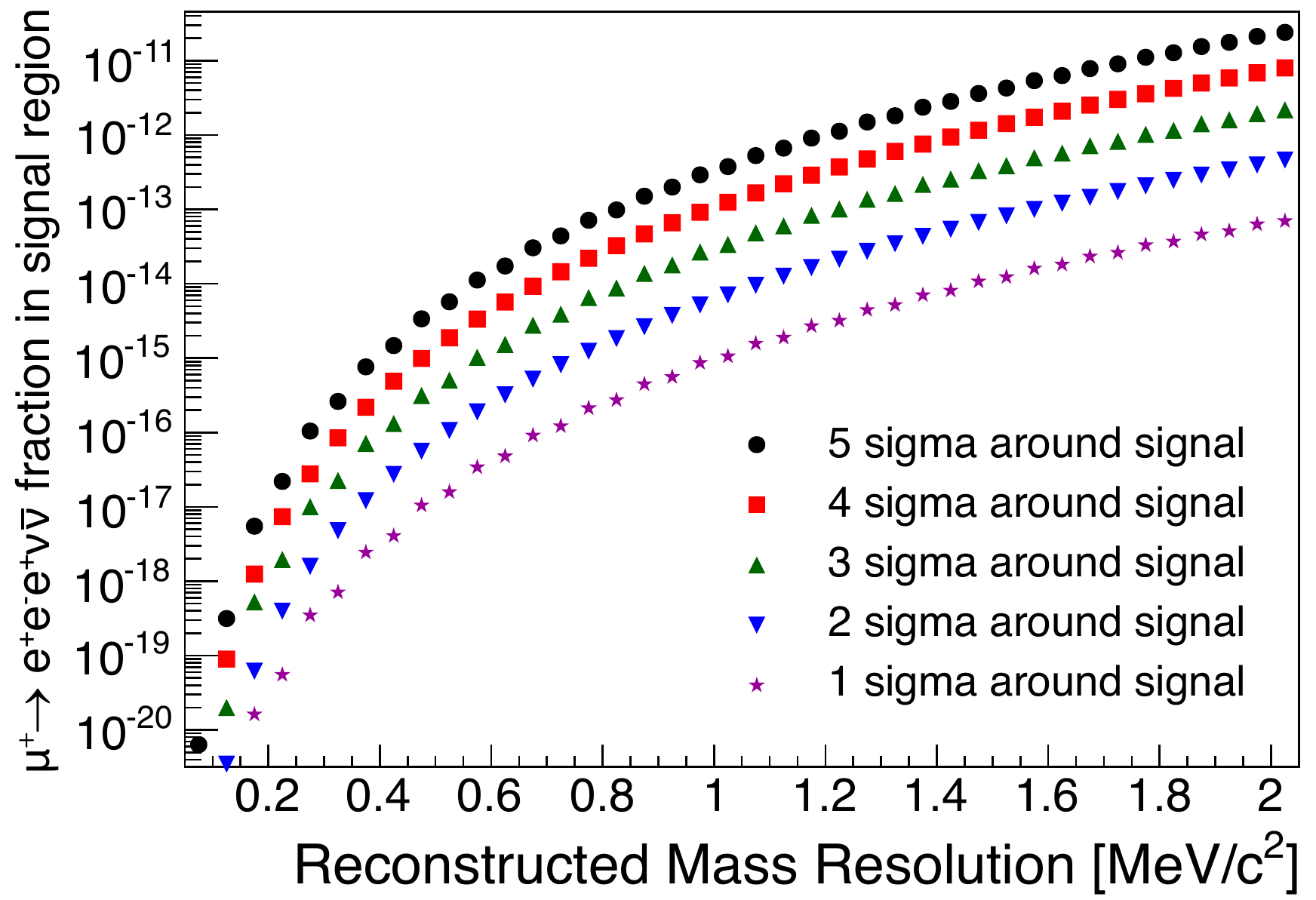}
\caption{Contamination of the signal region (one sided cut) with internal conversion (\mtenunu) events as a function of the visible three-particle mass resolution.
The branching fraction was taken from \cite{Djilkibaev:2008jy}, the resolution is assumed to be Gaussian.}
\label{fig:icontamination}
\end{figure}

The \emph{Mu3e} experiment \cite{RP} aims to find or exclude the decay \mte at the $10^{-16}$ level, improving the last measurement \cite{Bellgardt:1987du} by four orders of magnitude. 
Performing this measurement within a few years requires very intense, continuous muon beams, which are provided by the high-intensity proton accelerator at the Paul Scherrer Institute (PSI) in Switzerland. 
Muon stop rates in excess of $10^8$ per second are currently available. 
A future high-intensity muon beam (HIMB) capturing muons produced in the spallation target of the Swiss neutron source (SINQ) would push this beyond the $2 \cdot 10^{9}$ per second required to reach $10^{-16}$ in branching fraction sensitivity.
Running at rates of several billion muon decays per second and at the same time being able to suppress backgrounds by 16 orders of magnitude and being efficient for the signal is a formidable challenge for the detection system. 

\section{Requirements for the Detector}

\mte signal events are constituted by an electron and two positron tracks with a common vertex and coincident in time. 
As they originate from a muon decay at rest, the vectorial sum of their momenta should vanish and their energies should add up to the muon mass. 
The maximum momentum is $53$~MeV/$c$.
There are two main categories of background, on one hand accidental coincidences of positrons from ordinary muon decays with an electron from e.g.~photon conversion, Bhabha scattering or mis-reconstruction of positrons curling back in the magnetic field. 
Accidental backgrounds can be suppressed by excellent vertex, timing and momentum resolution. 
On the other hand there is the background due to the \emph{internal conversion} muon decay \mtenunu, which looks exactly like the signal decay, except that the neutrinos carry away some energy and momentum. 
From the branching fraction as a function of missing energy, the overall energy resolution required for a given sensitivity can be derived, see Fig.~\ref{fig:icontamination}.

\begin{figure}[t!]
	\centering
		\includegraphics[width=0.9\columnwidth,keepaspectratio]{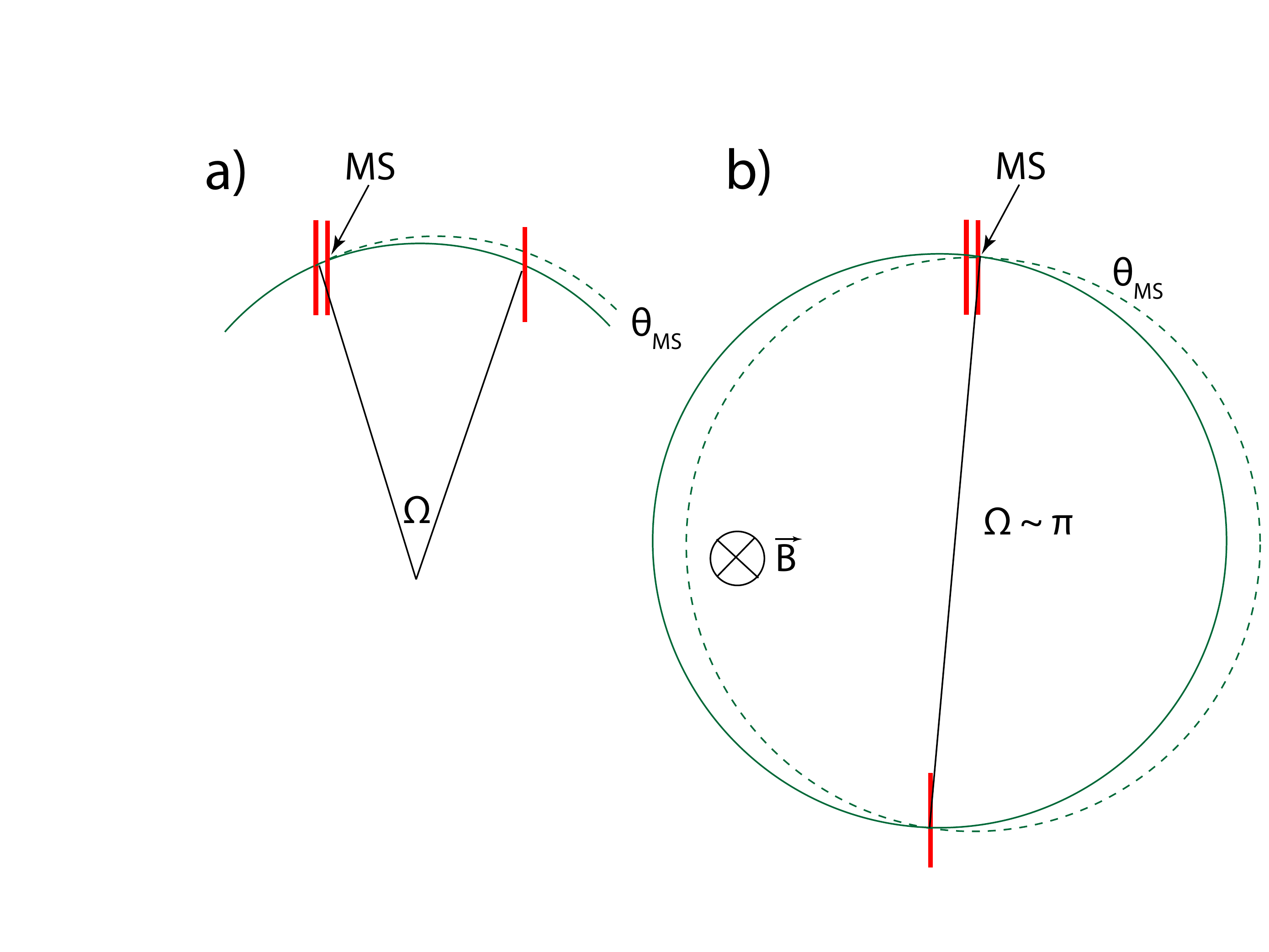}
	\caption{Multiple scattering and momentum measurement; a) for a short track segment, b) for a semi-circle.}
	\label{fig:MS}
\end{figure}

In summary, the Mu3e detector must provide excellent vertex and timing resolution as well as an average momentum resolution better than $0.5$~MeV/$c$ with a large geometrical acceptance and at the same time be capable of standing $2\cdot10^9$ muon decays per second.

A momentum measurement can be performed by measuring a direction (e.g.~with a double layer of detectors) and a third point with some lever arm to determine the curvature in the magnetic field. 
For the low momentum electrons\footnote{From here on meant to imply both electrons and positrons} in Mu3e, multiple Coulomb scattering (with expected scattering angle $\theta_{MS}$) and the lever arm $\Omega$ are the dominating quantities affecting the resolution, to first order
\begin{equation}
	\Delta p = \frac{\theta_{MS}}{\Omega},
\end{equation}
 see Fig.~\ref{fig:MS}a. 
For good momentum resolution it is thus imperative to have a minimum amount of material, which determines the detector technology as described in section \ref{sec:MAPS} and a large lever arm (ideally close to a semi-circle of the track, where multiple scattering effects cancel to first order, see Fig.~\ref{fig:MS}b), which determines the detector geometry, section \ref{sec:Det}.

\section{Detector Technology}
\label{sec:MAPS}

\begin{figure}[t!]
	\centering
		\includegraphics[width=0.8\columnwidth,keepaspectratio]{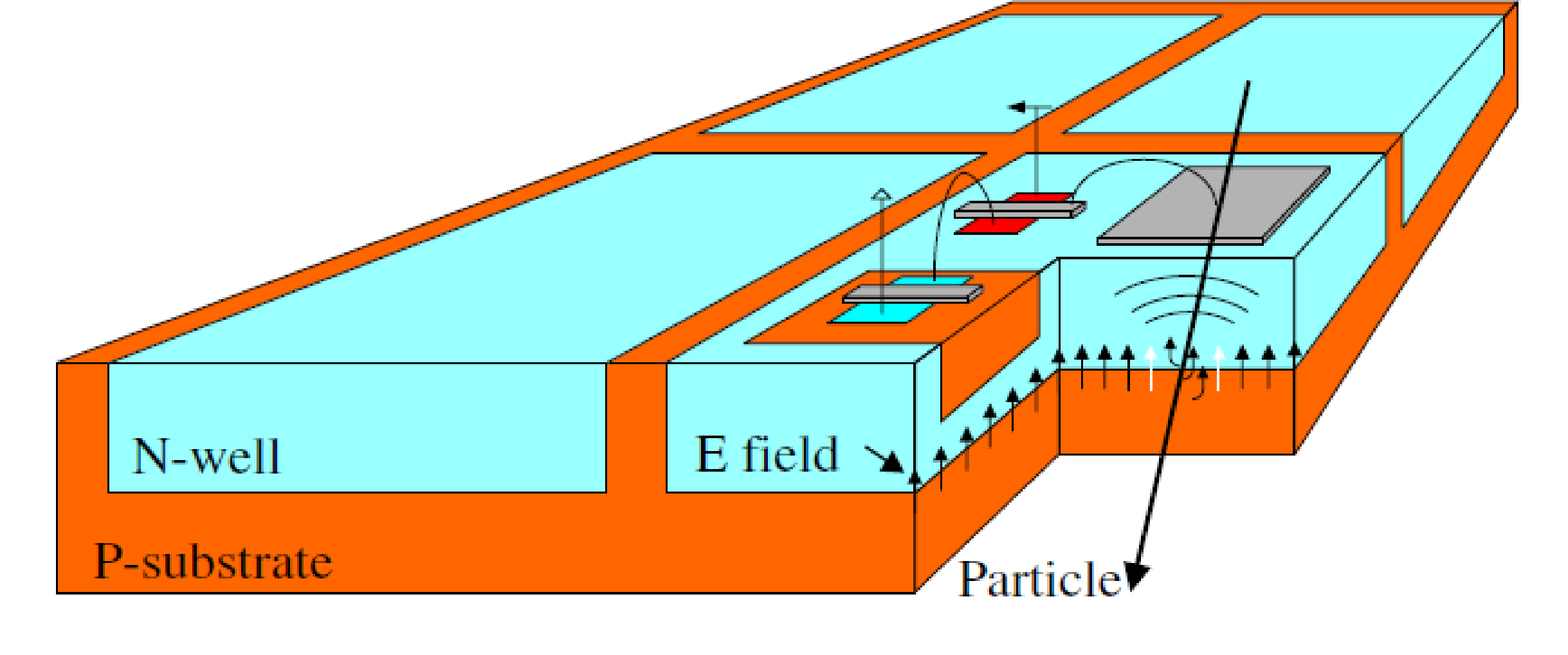}
	\caption{Schematic view of the HV-MAPS pixel sensor, from \cite{Peric:2007zz}.}
	\label{fig:HV_CMOS_Pixel_Sketch}
\end{figure}

\begin{figure}[b!]
	\centering
		\includegraphics[width=1.0\columnwidth,keepaspectratio]{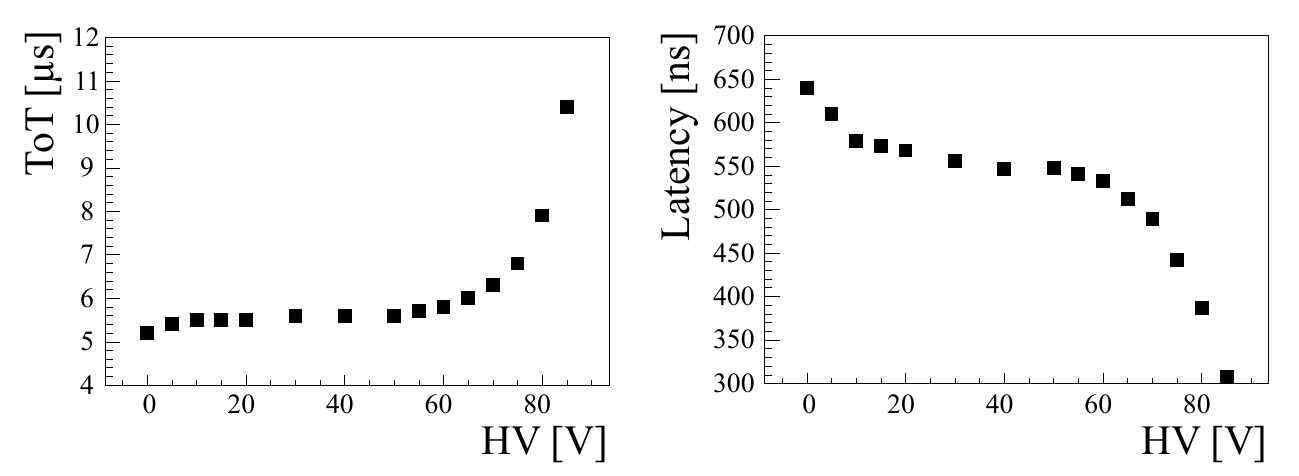}
	\caption{Time over threshold (signal size) and signal latency versus the applied high voltage in a MUPIX 2 prototype flashed with a LED, \cite{Perrevoort2012}.}
	\label{fig:tot_latency}
\end{figure}

For Mu3e, minimizing the material within the acceptance is crucial. 
Gas detectors however cannot stand the required rates due to ageing or occupancy (wire chambers) or do not deliver the required precision due to space charge effects (time-projection chambers, see e.g.~\cite{2012arXiv1209.0482B}). 
Solid state detectors until recently were either too thick (hybrid systems) or too slow (monolithic active pixel sensors, MAPS) for the task at hand. 
The high-voltage MAPS technology \cite{Peric:2007zz, Peric2010, Peric2010504} offers a way out of this dilemma by providing sensors that can be thinned to below 50~\textmu m thickness and ran at frame rates above 10~MHz due to fast charge collection and built-in zero-suppression. 

The MUPIX series of HV-MAPS are fabricated in a commercial process (AMS 180~nm HV-CMOS) and apply voltages of 50-100~V between the substrate and the deep $n$-wells containing the active electronics (see Fig.~\ref{fig:HV_CMOS_Pixel_Sketch} for a schematic drawing), leading to the fast charge collection via drift from a thin active depletion zone. 
The MUPIX 2 prototype with $42 \times 36$ pixels of $30 \times 39$ ~\textmu m$^2$ size was characterized extensively during 2012 \cite{Perrevoort2012, Augustin2012}. 
The chips implement a charge sensitive amplifier and a source follower inside the pixels and a comparator plus digital electronics in the chip periphery. 
The signal size is measured via the time-over-threshold method. 
For drift voltages above $\approx 70$~V signal amplification is observed, see Fig.~\ref{fig:tot_latency}, left. 
The use of LED flashes to generate signals allows for latency measurements between the flash and the appearance of the signal at the comparator exit. 
These latencies are of the order of 0.5~\textmu s, dominated by the shaping time and improving with high voltage, see Fig.~\ref{fig:tot_latency}, right.
The noise of the sensor was estimated using the sharpness of the threshold for a known injection signal and the signal size estimated using a $^{55}$Fe radioactive source, giving signal-to-noise-ratios well above 20.

The MUPIX 3 chip currently under study includes additional digital column logic and encodes hit information encompassing row and column address and a timestamp.
Also the pixels are now  $80 \times 92$~\textmu m$^2$ in size, close to the  $80 \times 80$~\textmu m$^2$ required for the production of $2 \times 1$ and $2 \times 2$~cm$^2$ sensors,
which will also include digital logic serializing the hit information on chip and sending it out via up to four 800~Mbit/s low-voltage differential signalling links.

\begin{figure}[t!]
	\centering
		\includegraphics[width=\columnwidth,keepaspectratio]{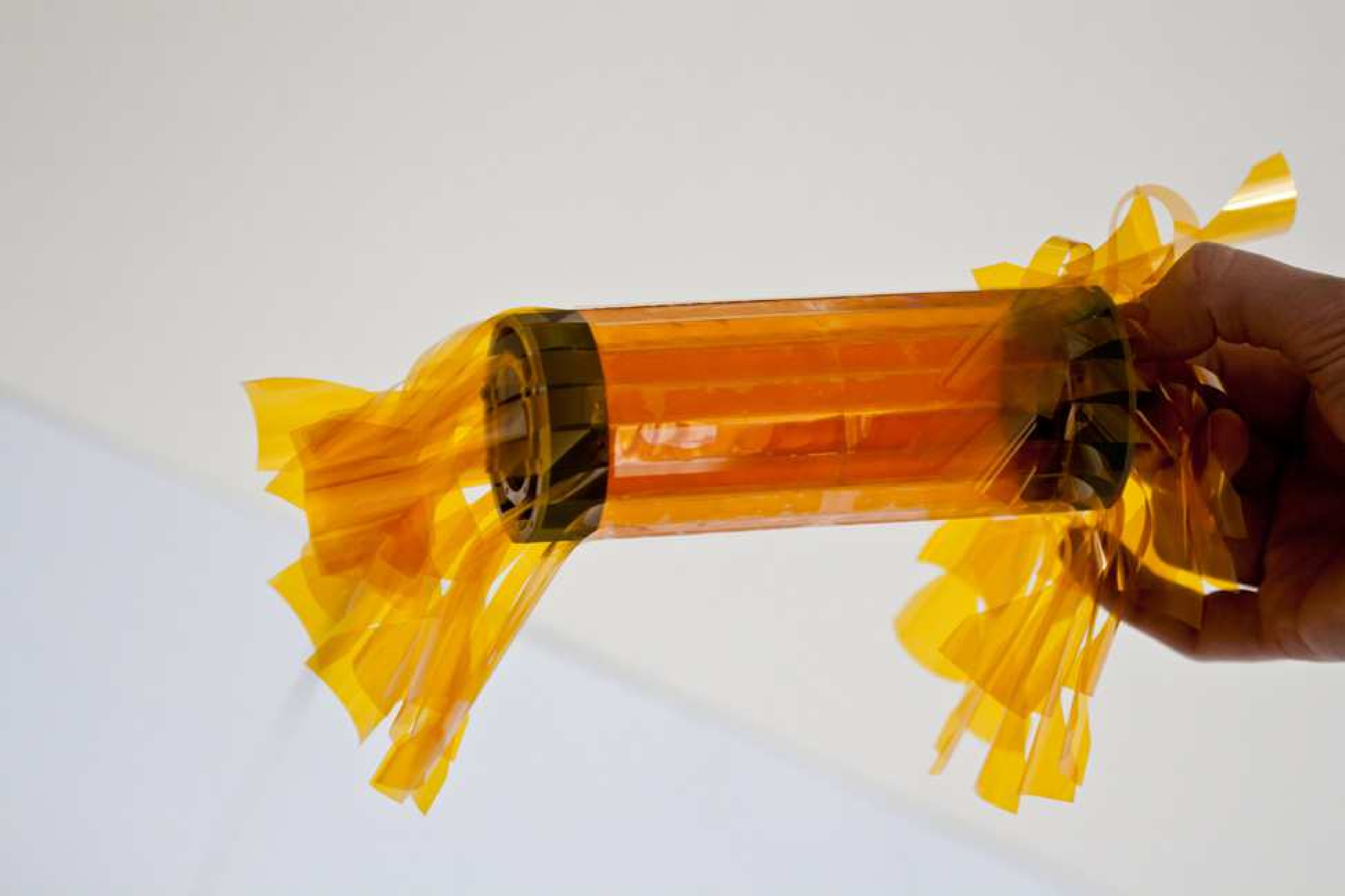}
	\caption{Prototype of the inner layer mechanical structure. The sensor chips are simulated by $100 \mu$m thin glass plates.}
	\label{fig:InnerLayerPrototypes}
\end{figure}

The MUPIX chips will be glued and bonded to Kapton\texttrademark~flex-prints with aluminium traces for signals and power.
These flex-prints in turn are glued to a Kapton\texttrademark~prism serving as a mechanical support.
A complete layer with  50~\textmu m silicon and twice 25~\textmu m Kapton\texttrademark~plus aluminium and adhesive is less than a permille of a radiation length thick.
Prototypes of this mechanical structure using glass plates instead of the silicon have been produced and were found to be surprisingly sturdy (and self-supporting), see Fig.~\ref{fig:InnerLayerPrototypes}.
The largest such structures required in the experiment cover 36~cm between supports. 
In order not to add dead material, the heating power of the chips (estimated at 150~mW/cm$^2$) will be cooled by a flow of gaseous Helium. 
How to prevent the necessary high flow rates of several m/s \cite{Zimmermann2012} from exciting vibrations in the sensor layers is currently under study.

\section{Detector Concept}
\label{sec:Det}

\begin{figure}[t!]
	\centering
		\includegraphics[width=\columnwidth,keepaspectratio]{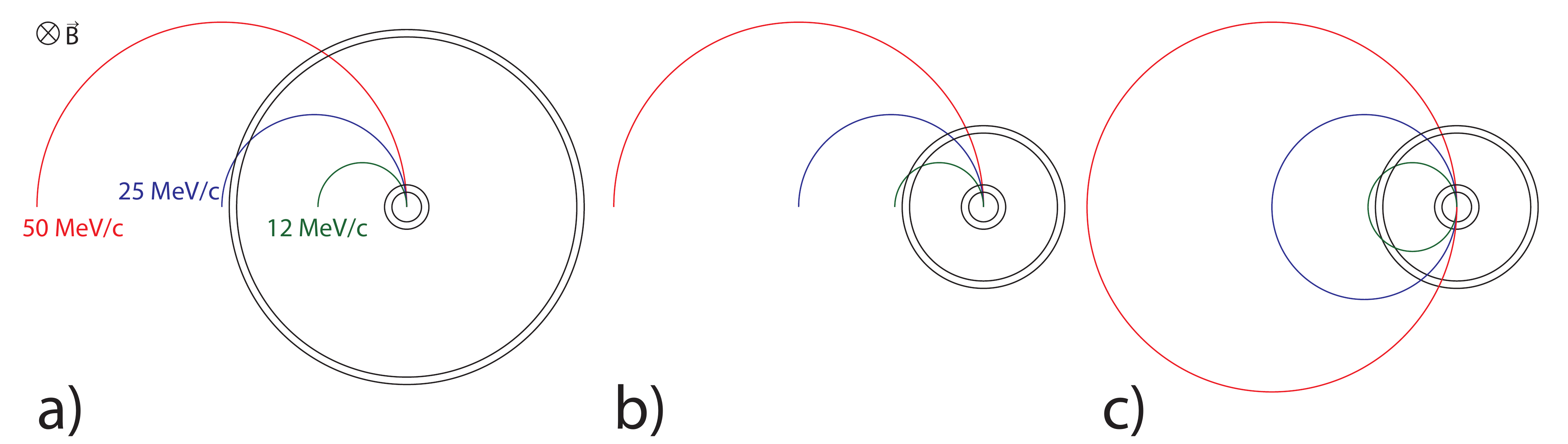}
	\caption{Tracks with transverse momenta of 12, 25 and 50~MeV/$c$ for a) a large radius detector optimized for precise measurements of large transverse momenta but with no acceptance at low momenta, b) a smaller detector optimized for acceptance at low momenta and c) the same small detector also measuring the re-curling parts of tracks, thus providing both high momentum resolution at large transverse momenta and good acceptance at low transverse momenta.}
	\label{fig:Momenta_Det}
\end{figure}

\begin{figure*}
	\centering
		\includegraphics[width=0.8\textwidth,keepaspectratio]{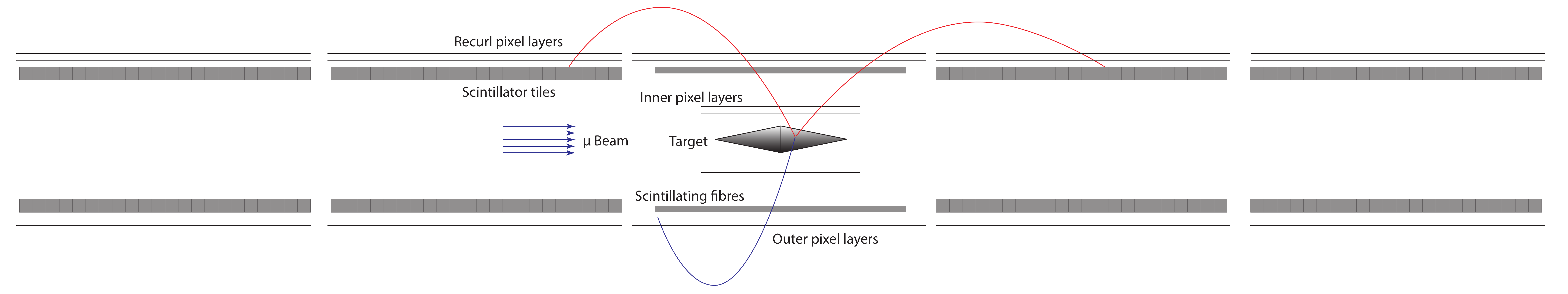}
	\caption{Schematic view of the detector design.}
	\label{fig:Schematic-9_notransverse}
\end{figure*}

\begin{figure}[t!]
	\centering
		\includegraphics[width=0.6\columnwidth,keepaspectratio]{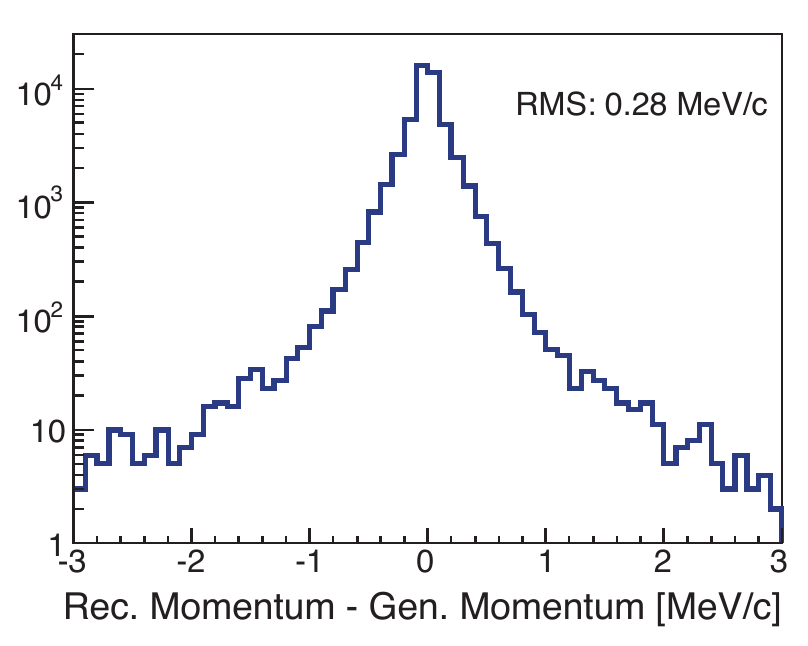}
	\caption{Momentum resolution for tracks following a Michel (muon decay) spectrum obtained with the multiple scattering only fit in the simplified model of the phase II detector.}
	\label{fig:MichelReso}
\end{figure}

\begin{figure}[t!]
	\centering
		\includegraphics[width=0.75\columnwidth,keepaspectratio]{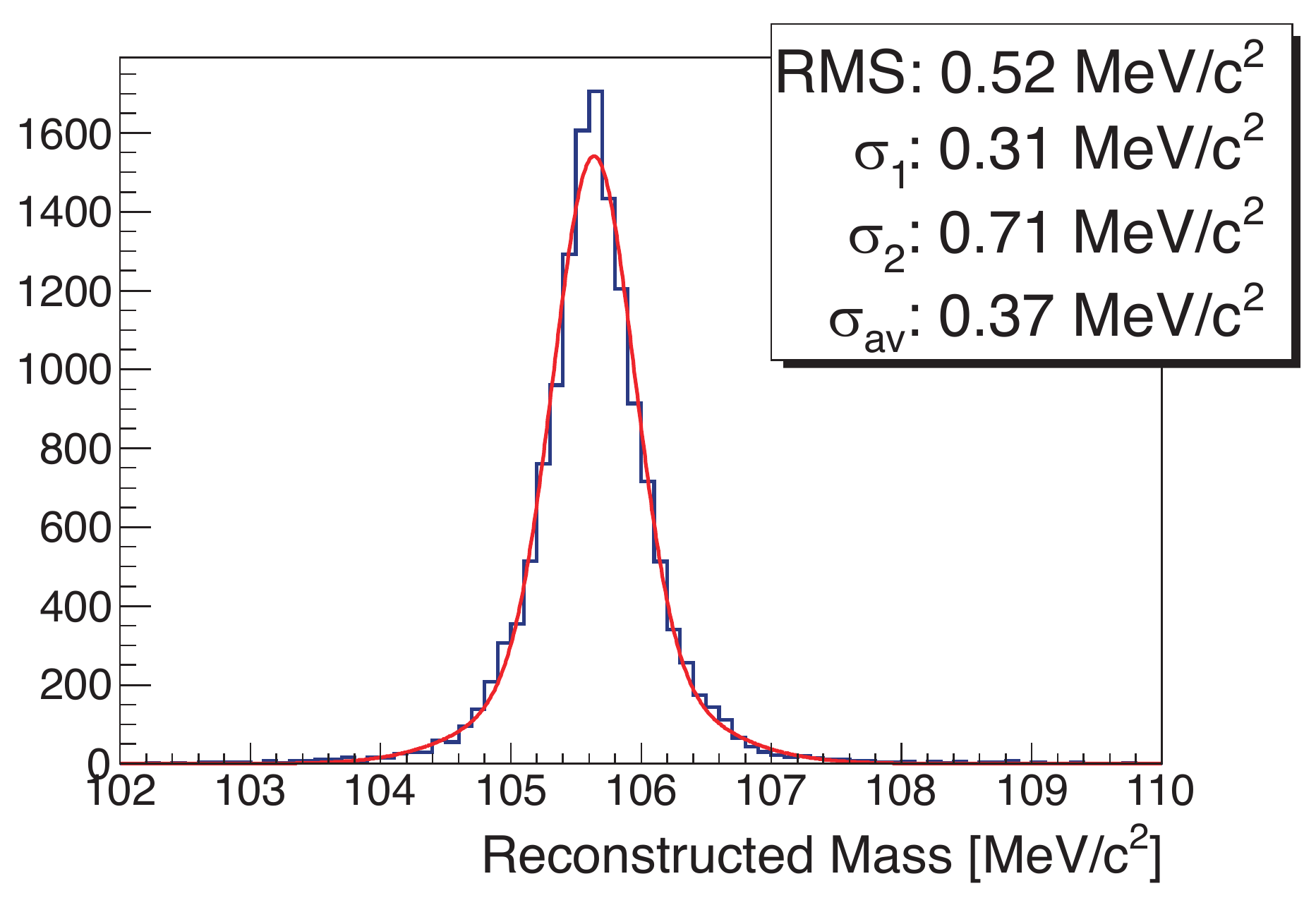}
	\caption{Reconstructed mass for signal decays in the simplified model of the phase II detector. The fit is with two Gaussian distributions.}
	\label{fig:MassReso}
\end{figure}

\begin{figure}
	\centering
		\includegraphics[width=0.85\columnwidth,keepaspectratio]{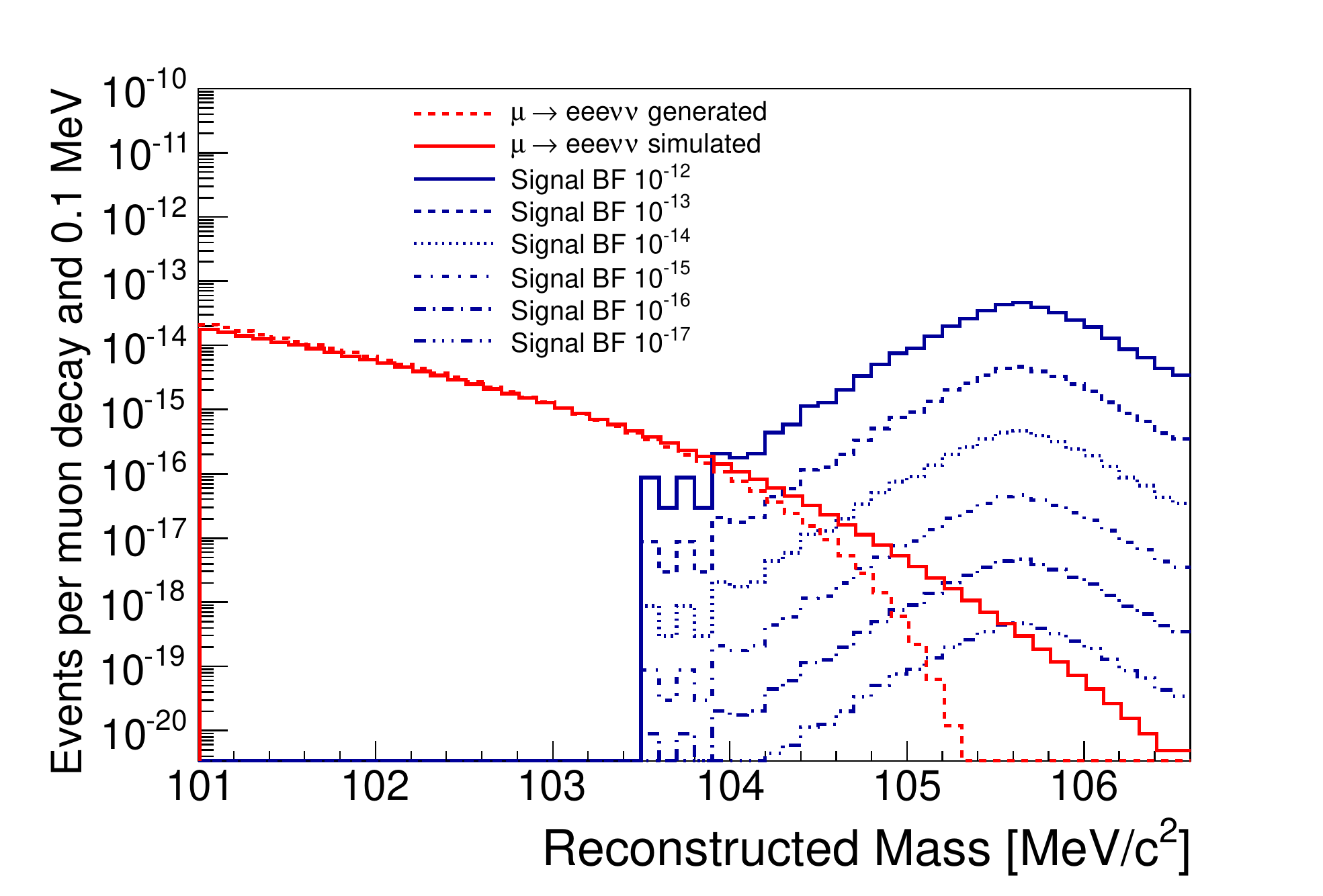}
	\caption{Tail of the internal conversion distribution overlaid with signal at different branching ratios for the phase II detector. The resolution for the internal conversion decays was taken from 30000 simulated signal decays.}
	\label{fig:Smearplot}
\end{figure}

The Mu3e tracking detector should be sensitive for a transverse momentum range from 12 to 53~MeV/$c$ and provide a momentum resolution of better than 0.5~MeV/$c$ over the same range. 
Moving the detectors to large radii in order to have a large lever arm for high transverse momentum tracks sacrifices acceptance at low momenta (Fig.~\ref{fig:Momenta_Det}a), whereas at small radii, the lever arm is not sufficient for delivering the required precision at large momenta (Fig.~\ref{fig:Momenta_Det}b). The Mu3e detector is small, but achieves a large lever arm, often close to the optimum of a semi-circle by also measuring the re-curling part of the track (Fig.~\ref{fig:Momenta_Det}c).
A good solid angle coverage for re-curling tracks leads to a long pipe design for the detector, see Fig.~\ref{fig:Schematic-9_notransverse}.
This design encompasses a hollow double-cone muon stopping target, made e.g.~from 30 to 80~\textmu m aluminium surrounded by two pixel detector layers for vertex determination. 
Between the inner and outer pixel detectors, three layers of 250~\textmu m scintillating fibres allow for a timing measurement with $\ord(1~\textrm{ns})$ resolution. 
The outer pixel layers and their extension in forward and backward direction allow for precise momentum measurement with re-curling tracks. 
Inside of the forward and backward extensions, scintillating tiles perform a timing measurement with a resolution better than 100~ps.
This overall arrangement of detectors leads to a minimum disturbance of the momentum (and vertex) measurement by the timing detectors, as the main contribution to the momentum resolution comes from the free space curl outside of the detector tube.
Another advantage of this design is the modularity, which allows for a staged approach in step with increasing muon rates.
A \emph{phase Ia} detector will consist of only the inner and outer central pixel detectors, sufficient for muon rates of a few 10$^7$/s. 
\emph{Phase Ib} will add the scintillating fibres and a first set of forward and backward extensions with pixels and tiles and run at $\approx$ 10$^8$ muons/s. 
The full detector as shown in Fig.~\ref{fig:Schematic-9_notransverse} together with a new beam line providing $2\times10^9$ muons/s will constitute the \emph{phase II} experiment. 

\section{Performance Studies}

The detector performance is studied both using a Geant4 \cite{Allison:2006ve, Agostinelli2003250} based simulation with a detailed description of materials and the magnetic field as well as a simplified model mainly for reconstruction studies assuming perfectly cylindrical tracking layers.
A three-dimensional multiple scattering track fit ignoring detector resolution is employed for track finding.
It is planned to employ a general broken line fit \cite{Blobel:2011az,Kleinwort:2012yt,Kiehn2012} in a second step in order to obtain the best possible resolution.
The results presented in the following were obtained with a simplified model and a multiple scattering dominated fit.
Fig.~\ref{fig:MichelReso} shows the momentum resolution for tracks from ordinary muon decays with an RMS of 280~keV/$c$.
The resolution for the invariant mass of the three decay particles is shown in Fig.~\ref{fig:MassReso}. Fig.~\ref{fig:Smearplot} finally shows the ability of the presented detection system for separating the \mte signal from the \mtenunu background.

\section{Conclusions}

The Mu3e tracking detector employs a variety of innovative techniques in order to obtain the best possible momentum resolution in a multiple scattering dominated environment, namely:
\begin{itemize}
 \itemsep0em
	\item HV-MAPS sensors thinned to 50~\textmu m thickness;
	\item A mechanical structure built from 25~\textmu m Kapton\texttrademark~foil and a 25~\textmu m Kapton\texttrademark~flexprint, leading to a layer thickness of less than a permille of a radiation length;
	\item Cooling with gaseous helium in order not to add material;
	\item The use of recurling tracks for increasing the lever arm for momentum measurements.
\end{itemize}
The combination of these techniques achieves a three particle invariant mass resolution in the order of 0.5~MeV/$c^2$, thus fulfilling the requirements of the Mu3e experiment for reaching $10^{-16}$ in branching ratio sensitivity for the decay \mte.

\section*{Acknowlegments}

N.~Berger would like to thank the \emph{Deutsche Forschungsgemeinschaft} for supporting him and the Mu3e project through an Emmy Noether grant. M.~Kiehn acknowledges support by the IMPRS-PTFS.





\end{document}